\documentclass[a4paper]{article}
\usepackage[margin=1.6in]{geometry}

\usepackage[english]{babel}
\usepackage[utf8]{inputenc}
\usepackage{amsmath}
\usepackage{amsthm}
\usepackage{amssymb}
\usepackage{mathtools}
\usepackage{graphicx}
\usepackage{url}
\usepackage[numbers]{natbib}
\usepackage{epstopdf}
\usepackage{textcomp}

\usepackage{tikz}
\usetikzlibrary{fit,shapes,arrows}

\tikzstyle{block} = [draw, rectangle, minimum height=3em, minimum width=3em]
\tikzstyle{sum} = [draw, circle, node distance=1cm]
\tikzstyle{input} = [coordinate]
\tikzstyle{output} = [coordinate]
\tikzstyle{pinstyle} = [pin edge={to-,thin,black}]

\renewcommand{\S}{\textsection}

\renewcommand{\vec}{\mathbf}
\newcommand{\dotp}[3]{\left\langle {#1}, {#2} \right\rangle_{#3}}

\newtheorem{theorem}{Theorem}[section]

\title{Methods for applying the
Neural Engineering Framework
to neuromorphic hardware}

\author{Aaron R. Voelker and Chris Eliasmith \\
Centre for Theoretical Neuroscience, University of Waterloo}

\date{\today}

\begin{document}
\maketitle

\begin{abstract}
We review our current software tools and theoretical methods for applying the Neural Engineering Framework to state-of-the-art neuromorphic hardware.
These methods can be used to implement linear and nonlinear dynamical systems that exploit axonal transmission time-delays, and to fully account for nonideal mixed-analog-digital synapses that exhibit higher-order dynamics with heterogeneous time-constants.
This summarizes earlier versions of these methods that have been discussed in a more biological context (Voelker~\&~Eliasmith,~2017) or regarding a specific neuromorphic architecture (Voelker et al.,~2017).
\end{abstract}

\section{Introduction}

This report is motivated by the recent influx of neuromorphic computing architectures including SpiNNaker~\citep{mundy2015efficient}, Neurogrid~\citep{choudhary2012silicon}, Brainstorm~\citep{brainstorm}, IBM's TrueNorth~\citep{merolla2014million}, BrainScaleS~\citep{schemmel2010wafer}, and ROLLS~\citep{qiao2015reconfigurable}, among others.
These are massively-parallel, low-power, analog, and/or digital systems (often a mix of the two) that are designed to simulate large-scale artificial neural networks rivalling the scale and complexity of real biological systems.
Despite growing excitement, we believe that methods to fully unlock the computational power of neuromorphic hardware are lacking. 
This is primarily due to a theoretical gap between our traditional, discrete, {\it von Neumann}-like understanding of conventional algorithms and the continuous spike-based signal processing of real brains that is often emulated in silicon~\citep{boahen2017neuromorph}.

We use the term ``neural compiler'' to refer loosely to any systematic method of converting an algorithm, expressed in some high-level mathematical language, into synaptic connection weights between populations of spiking neurons~\citep{bekolay2014}.
To fully leverage neuromorphic hardware for real-world applications, we require neural compilers that can account for the effects of spiking neuron models and mixed-analog-digital synapse models, and, perhaps more importantly, exploit these details in useful ways when possible.
There exist various approaches to neural engineering, including 
those by Den{\`e}ve~et~al.~\citep{boerlin2013predictive, schwemmer2015constructing} and Memmesheimer~et~al.~\citep{thalmeier2016learning}.
However, the Neural Engineering Framework~(NEF;~\citep{eliasmith1999developing, eliasmith2003neural}) stands apart in terms of software implementation~(Nengo;~\citep{bekolay2014, stewart2009a, Sharma2016, gosmann2017}), large-scale cognitive modeling~\citep{eliasmith2012large, eliasmith2013build}, and neuromorphic applications~\citep{mundy2015efficient, choudhary2012silicon, brainstorm, dethier2011, corradi2014, Knight2016, berzish2016, mundyreal}.
Competing methods consistently exclude important details, such as refractory periods and membrane voltage leaks, from their networks~\citep{boerlin2013predictive, thalmeier2016learning}.
The NEF, on the other hand, embraces biological complexity whenever it proves computationally useful~\cite{voelker2017b} and/or improves contact with neuroscience literature~\cite{eliasmith2016}.
This approach to biological modeling directly mirrors a similar need to account for the details in neuromorphic hardware when building neural networks~\citep{dethier2011, corradi2014, voelker2017a}.

Nevertheless, there are many open problems in optimizing the NEF for state-of-the-art neuromorphics.
In particular, we have been working to account for more detailed dynamics in heterogeneous models of the post-synaptic current (PSC) induced by each spike, as well as delays in spike propagation~\citep{voelker2017b, voelker2017a, dynamicspatent, nengolib}.
The purpose of this report is to summarize our methods, both theoretical and practical, that have progressed in this direction.
There are similar challenges in extending the NEF to account for multi-compartment neuron models~\citep{eliasmith2016, duggins2017}, conductance-based synapses~\citep{stockel2017}, and to minimize the total number of spikes~\citep{boahen2017neuromorph} -- but these topics will not be addressed in this report.

The remainder of this report assumes that the reader is already comfortable with Nengo and the NEF in the context of engineering dynamical systems~(i.e.,~``Principle~III''). 
A technical yet accessible overview of the NEF can be found in \citep{stewart2012technical}, and a tutorial for Nengo~(version 2) can be found in \citep{Sharma2016}.

\section{Accounting for Synaptic Dynamics}
\label{sec:principle3}

This section provides a theoretical account of the effect of higher-order linear synapse models on the dynamics of the network, by summarizing the extensions from \citep{voelker2017b} and \citep{voelker2017a}.
This yields two novel proofs of Principle~III from the NEF, and generalizes the principle to include more detailed synapses, including those modeling axonal transmission delays.

\subsection{Linear systems}
\label{sec:lti}

Here we focus our attention on linear time-invariant (LTI) systems:
\begin{equation} \label{eq:lti}
\begin{aligned}
\dot{\vec{x}}(t) &= A\vec{x}(t) + B\vec{u}(t) \\
\vec{y}(t) &= C\vec{x}(t) + D\vec{u}(t)
\end{aligned}
\end{equation}
where the time-varying vector $\vec{x}(t)$ represents the system state, $\vec{y}(t)$ the output, $\vec{u}(t)$ the input, and the time-invariant ``state-space'' matrices $(A\text{,}\, B\text{,}\, C\text{,}\, D)$ fully determine the system's dynamics.
We will omit the variable $t$ when not needed.

Principle~III from the NEF states that in order to train the recurrent connection weights to implement (\ref{eq:lti})---using a continuous-time lowpass filter \mbox{$h(t) = (1 / \tau) e^{-t/\tau}$} to model the PSC---we use Principle~II to train the decoders for the recurrent transformation $(\tau A + I)\vec{x}$, input transformation $\tau B \vec{u}$, output transformation $C\vec{x}$, and passthrough transformation $D\vec{u}$~\citep[][pp.~\mbox{221--225}]{eliasmith2003neural}.
This drives the recurrent synapses with the signal $\tau \dot{\vec{x}} + \vec{x}$ so that their output is the signal $\vec{x}$, in effect transforming the synapses into perfect integrators with respect to $\dot{\vec{x}}$.
The vector $\vec{x}$ is then represented by the population of neurons via Principle~I.
Thus, this provides a systematic approach for training a recurrent neural network to implement any linear dynamical system.
Now we show that this approach generalizes to other synaptic models.

For these purposes, the transfer function is a more useful description of the LTI system than (\ref{eq:lti}).  
The transfer function is defined as the ratio of $Y(s)$ to $U(s)$, given by the Laplace transforms of $\vec{y}(t)$ and $\vec{u}(t)$ respectively. 
The variable $s$ denotes a complex value in the frequency domain, while $t$ is non-negative in the time domain. 
The transfer function is related to (\ref{eq:lti}) by the following:
\begin{align} \label{eq:tf}
F(s) = \frac{Y(s)}{U(s)} = C(sI - A)^{-1}B + D \text{.}
\end{align}

The transfer function $F(s)$ can be converted into the state-space model $(A\text{,}\, B\text{,}\, C\text{,}\, D)$ if and only if it can be written as a proper ratio of finite polynomials in $s$. 
The ratio is proper when the degree of the numerator does not exceed that of the denominator. 
In this case, the output will not depend on future input, and so the system is `causal'. 
The order of the denominator corresponds to the dimensionality of $\vec{x}$, and therefore must be finite. 
Both of these conditions can be interpreted as physically realistic constraints where time may only progress forward, and neural resources are finite.

\paragraph{Continuous Identity}

In order to account for the introduction of a synaptic filter $h(t)$, we replace the integrator $s^{-1}$ in (\ref{eq:tf}) with $H(s)$, where $H(s) = \mathcal{L} \left\{ h(t) \right\}$.
This new system has the transfer function $C(H(s)^{-1}I - A)^{-1}B + D = F(H(s)^{-1})$.
To compensate for this change in dynamics, we must invert the change-of-variables $s \leftrightarrow H(s)^{-1}$.
This means finding the {\it required} $F^H(s)$ such that $F^H(H(s)^{-1})$ is equal to the {\it desired} transfer function, $F(s)$.
We highlight this as the following identity:
\begin{align} \label{eq:tf-identity}
\Aboxed{ F^H \left( \frac{1}{H(s)} \right) = F(s) \text{.} } 
\end{align}
Then the state-space model $(A^H\text{,}\, B^H\text{,}\, C^H\text{,}\, D^H)$ satisfying (\ref{eq:tf}) with respect to $F^H(s)$ will implement the desired dynamics (\ref{eq:lti}) given $h(t)$.

\paragraph{Discrete Identity}

For the discrete (i.e.,~digital~synapse) case, we begin with $F(z)$ and $H(z)$ expressed as digital systems.
The form of $H(z)$ is usually determined by the hardware, and $F(z)$ is usually found by a zero-order hold (ZOH) discretization of $F(s)$ using the simulation time-step ($dt$), resulting in the discrete LTI system:
\begin{equation} \label{eq:lti-discrete}
\begin{aligned}
\vec{x}[t+dt] &= \bar{A}\vec{x}[t] + \bar{B}\vec{u}[t] \\
\vec{y}[t] &= \bar{C}\vec{x}[t] + \bar{D}\vec{u}[t] \text{.}
\end{aligned}
\end{equation}
Here, we have the same relationship as (\ref{eq:tf}),
\begin{align} \label{eq:tf-discrete}
F(z) = \frac{Y(z)}{U(z)} = \bar{C}(zI - \bar{A})^{-1}\bar{B} + \bar{D} \text{.}
\end{align}
Therefore, the previous discussion applies, and we must find an $F^H(z)$ that satisfies:
\begin{align} \label{eq:tf-identity-discrete}
\Aboxed{ F^H \left( \frac{1}{H(z)} \right) = F(z) \text{.} } 
\end{align}
Then the state-space model $(\bar{A}^H\text{,}\, \bar{B}^H\text{,}\, \bar{C}^H\text{,}\, \bar{D}^H)$ satisfying (\ref{eq:tf-discrete}) with respect to $F^H(z)$ will implement the desired dynamics (\ref{eq:lti-discrete}) given $h[t]$.
In either case, the general problem reduces to solving this change-of-variables problem for various synaptic models.
We now provide a number of results.
More detailed derivations are available in \citep{voelker2017b}.

\paragraph{Continuous Lowpass Synapse}

Replacing the integrator $s^{-1}$ with the standard continuous-time lowpass filter, so that $H(s) =  \frac{1}{\tau s + 1}$:
\begin{align} \label{eq:p3cont}
F^H(\tau s + 1) = F(s) \iff F^H(s) = C(sI - (\tau A + I))^{-1}(\tau B) + D
\end{align}
which rederives the standard form of Principle~III from the NEF~\citep{eliasmith2003neural}.

\paragraph{Discrete Lowpass Synapse}

Replacing the integrator with a discrete-time lowpass filter $H(z) = \frac{1 - a}{z - a}$ in the $z$-domain with time-step $dt$, where $a = e^{-\frac{dt}{\tau}}$:
\begin{align} \label{eq:p3discrete}
F^H \left( \frac{z-a}{1-a} \right) = F(z) \iff F^H(z) = \bar{C} \left( zI - \frac{1}{1 - a}(\bar{A} - aI) \right)^{-1} \left( \frac{1}{1-a}\bar{B} \right) + \bar{D} \text{.}
\end{align}
Therefore, $\bar{A}^H = \frac{1}{1 - a}(\bar{A} - aI)$, $\bar{B}^H = \frac{1}{1-a}\bar{B}$, $\bar{C}^H = \bar{C}$, and $\bar{D}^H = \bar{D}$.
This mapping can dramatically improve the accuracy of Principle~III in digital simulations (e.g.,~when using a desktop computer)~\citep{voelker2017b}.

\paragraph{Delayed Continuous Lowpass Synapse}

Replacing the integrator with a continuous lowpass filter containing a pure time-delay of length $\lambda$, so that $H(s) = \frac{e^{-\lambda s}}{\tau s + 1}$:
\begin{align}
F^H \left( \frac{\tau s + 1}{e^{-\lambda s}} \right) = F(s) \iff F^H(s) = F \left( \frac{1}{\lambda} W_0(ds) - \frac{1}{\tau} \right) \text{,}
\end{align}
where $d = \frac{\lambda}{\tau}e^{\frac{\lambda}{\tau}}$ and $W_0(xe^x) = x$ is the principal branch of the Lambert-$W$ function~\citep{corless1996lambertw}.\footnote{This assumes that $|\eta| < \pi$ and $\lambda \text{Re}\left[ s \right] + \frac{\lambda}{\tau} > - \eta \cot \eta$, where $\eta := \lambda \text{Im}\left[ s \right]$~\citep{voelker2017b}.}
This synapse model can be used to model axonal transmission time-delays due to the finite-velocity propagation of action potentials, or to model feedback delays within a broader control-theoretic context. 
To demonstrate the case where a pure time-delay of length $\theta$ is the desired transfer function ($F(s) = e^{-\theta s}$), we let $c = e^{\frac{\theta}{\tau}}$ and $r = \frac{\theta}{\lambda}$ to obtain the required transfer function:
\begin{align} \label{eq:lowdelay}
F^H(s) = c \left( \frac{W_0(ds)}{ds} \right)^r = cr \sum_{i=0}^\infty \frac{(i+r)^{i-1}}{i!} (-ds)^i \text{.}
\end{align}
We then numerically find the Pad\'e approximants of the latter Taylor series.
More details and validation may be found in \citep{voelker2017b} and \citep{dynamicspatent}.

\paragraph{General Case}

Finally, we consider any linear synapse model of the form:
\begin{equation} \label{eq:synapse}
H(s) = \frac{1}{\sum_{i=0}^k c_i s^i} \text{,}
\end{equation}
for some polynomial coefficients $\left( c_i \right)$ of arbitrary degree $k$.
To the best of our knowledge, this class includes the majority of linear synapse models used in the literature.
For synapses containing a polynomial numerator with degree $q > 1$ (e.g., considering the Taylor series expansion of the box filter $\epsilon^{-1} (1 - e^{-\epsilon s}) s^{-1}$), we take its $[0\text{,}~q]$-Pad\'e approximants to transform the synapse into this form within some radius of convergence.\footnote{This is equivalent to the approach taken in \citep[][equations (9)-(11)]{voelker2017a}.}
To map $F(s)$ onto (\ref{eq:synapse}), we begin by defining our solution to $F^{H}(s)$ in the form of its state-space model:
\begin{equation} \label{eq:general-linear}
\begin{aligned}
A^H &= \sum_{i=0}^k c_i A^i \text{,} & \quad C^H &= C \text{,} \\
B^H &= \left( \sum_{j=0}^{k-1} s^j \sum_{i=j+1}^k c_i A^{i-j-1} \right) B \text{,} & \quad D^H &= D \text{,}
\end{aligned}
\end{equation}
which we claim satisfies (\ref{eq:tf-identity}).
To validate this claim, we assert the following algebraic relationship (proven in \citep{voelker2017b}):
\begin{align*}
H(s)^{-1}I - A^H = \sum_{i=0}^k c_i \left( s^i I - A^i \right) = \left( \sum_{j=0}^{k-1} s^j \sum_{i=j+1}^k c_i A^{i-j-1} \right) (sI - A)\text{.}
\end{align*}
We now verify that (\ref{eq:general-linear}) satisfies (\ref{eq:tf-identity}):
\begin{align*}
F^H(H(s)^{-1}) &= C^H(H(s)^{-1}I - A^H)^{-1} B^H + D^H \\
&= C(H(s)^{-1}I - A^H)^{-1} \left( \sum_{j=0}^{k-1} s^j \sum_{i=j+1}^k c_i A^{i-j-1} \right) B + D \\
&= C(sI - A)^{-1} B + D = F(s) \text{.} \tag*{\qed}
\end{align*}
Since $s^j$ is the $j^{\text{th}}$-order differential operator, this form of $B^H$ states that we must supply the $j^{\text{th}}$-order input derivatives $\vec{u}^{(j)}$, for all $j = 1 \ldots k - 1$.
To be more precise, let us first define $B^H_j:= \left( \sum_{i=j+1}^k c_i A^{i-j-1} \right) B$.
Then (\ref{eq:general-linear}) states that the ideal state-space model must implement the input transformation as a linear combination of input derivatives, $\sum_{j=0}^{k-1} B^H_j \vec{u}^{(j)}$.
However, if the required derivatives are not included in the neural representation, then it is natural to use a ZOH method by assuming $\vec{u}^{(j)} = 0$, for all $j = 1 \ldots k - 1$:
\begin{equation} \label{eq:general-linear-approx}
B^H = \left( \sum_{i=1}^k c_i A^{i-1} \right) B \text{,}
\end{equation}
with $A^H$, $C^H$, and $D^H$ as in (\ref{eq:general-linear}).
This is now an equivalent model to (\ref{eq:general-linear}) assuming ZOH, and in the form of the standard state-space model (\ref{eq:lti}).

The same derivation also applies to the discrete-time domain, with respect to the discrete synapse (corresponding to some implementation in digital hardware):
\begin{equation} \label{eq:synapse-discrete}
H(z) = \frac{1}{\sum_{i=0}^k \bar{c}_i z^i} \text{.}
\end{equation}
Here, the only real difference (apart from notation) is the discrete version of (\ref{eq:general-linear-approx}):
\begin{equation} \label{eq:general-linear-approx-discrete}
\bar{B}^H = \left( \sum_{j=0}^{k-1} \sum_{i=j+1}^k \bar{c}_i \bar{A}^{i-j-1} \right) \bar{B} \text{.}
\end{equation}
These mappings are made available by \texttt{ss2sim} and \texttt{LinearNetwork} in nengolib~0.4.0~\citep{nengolib}.
Additional details may again be found in \citep{voelker2017b}.

\subsection{Nonlinear systems}
\label{sec:nonlinear}

Here we derive two theorems for nonlinear systems, by taking a different perspective that is consistent with \S\ref{sec:lti}.
This generalizes the approach taken in \citep{voelker2017a}, which considered the special case of a pulse-extended (i.e.,~time-delayed) double-exponential.

We wish to implement some desired nonlinear dynamical system,
\begin{equation} \label{eq:nonlinear}
\dot{\vec{x}}(t) = f(\vec{x}, \vec{u}) \text{,}
\end{equation}
using (\ref{eq:synapse}) as the synaptic filter $h(t)$.
Letting $\vec{w}(t) = f^h(\vec{x}, \vec{u})$ for some recurrent function $f^h$ and observing that $\vec{x}(t) = (\vec{w} \ast h)(t)$, we may express these dynamics in the Laplace domain:
\begin{align*}
&& \frac{\vec{X}(s)}{\vec{W}(s)} &= \frac{1}{\sum_{i=0}^k c_i s^i} \\
\iff && \vec{W}(s) &= \vec{X}(s) \sum_{i=0}^k c_i s^i = \sum_{i=0}^k c_i \left[ s^i \vec{X}(s) \right] \\
\iff && \vec{w}(t) &= \sum_{i=0}^k c_i \vec{x}^{(i)}
\end{align*}
since $s$ is the differential operator. This proves the following theorem:
\begin{theorem} \label{thm:p3cont-nonlinear}
Let the function computed along the recurrent connection be:
\begin{align}
\Aboxed{f^h(\vec{x}, \vec{u}) = \sum_{i=0}^k c_i \vec{x}^{(i)}}
\end{align}
where $\vec{x}^{(i)}$ denotes the $i^\text{th}$ time-derivative of $\vec{x}(t)$, and $c_i$ are given by (\ref{eq:synapse}). Then the resulting dynamical system is precisely (\ref{eq:nonlinear}).
\end{theorem}

For the discrete case, we begin with some desired nonlinear dynamics expressed over discrete time-steps:
\begin{equation} \label{eq:disc-nonlinear}
\vec{x}[t+dt] = \bar{f}(\vec{x}, \vec{u}) \text{,}
\end{equation}
using (\ref{eq:synapse-discrete}) as the synaptic filter $h[t]$, followed by an analogous theorem:
\begin{theorem} \label{thm:p3disc-nonlinear}
Let the function computed along the recurrent connection be:
\begin{align}
\Aboxed{\bar{f}^h(\vec{x}, \vec{u}) = \sum_{i=0}^k \bar{c}_i \vec{x}^{[i]}}
\end{align}
where $\vec{x}^{[i]}$ denotes the $i^\text{th}$ discrete forwards time-shift of $\vec{x}$, and $\bar{c}_i$ are given by (\ref{eq:synapse-discrete}).
Then the resulting dynamical system is precisely (\ref{eq:disc-nonlinear}).
\end{theorem}
The proof for the discrete case is nearly identical.
For sake of completeness, let $\vec{w}[t] = \bar{f}^h(\vec{x}, \vec{u})$ for some recurrent function $\bar{f}^h$ and observe that $\vec{x}[t] = (\vec{w} \ast h)[t]$:
\begin{align*}
&& \frac{\vec{X}(z)}{\vec{W}(z)} &= \frac{1}{\sum_{i=0}^k \bar{c}_i z^i} \\
\iff && \vec{W}(z) &= \vec{X}(z) \sum_{i=0}^k \bar{c}_i z^i = \sum_{i=0}^k \bar{c}_i \left[ z^i \vec{X}(z) \right] \\
\iff && \vec{w}[t] &= \sum_{i=0}^k \bar{c}_i \vec{x}^{[i]}
\end{align*}
since $z$ is the forwards time-shift operator. \qed

\paragraph{Continuous Lowpass Synapse}

For standard Principle~III, we have $H(s) = \frac{1}{\tau s + 1}$ $\implies$ $k = 1$, $c_0 = 1$ and $c_1 = \tau$, 
\begin{equation} \label{eq:p3cont-nonlinear}
\implies f^h(\vec{x}, \vec{u}) = c_0 \vec{x}^{(0)} + c_1 \vec{x}^{(1)} = \vec{x} + \tau \dot{\vec{x}} = \tau f(\vec{x}, \vec{u}) + \vec{x} \text{.}
\end{equation}
Note that (\ref{eq:p3cont-nonlinear}) is consistent with (\ref{eq:p3cont}) and with Principle~III from the NEF.

\paragraph{Discrete Lowpass Synapse}

For the discrete case of Principle~III, we have $H(z) = \frac{1 - a}{z - a}$, where $a = e^{-dt / \tau}$ $\implies$ $k = 1$, $\bar{c}_0 = -a(1 - a)^{-1}$, $\bar{c}_1 = (1 - a)^{-1}$,
\begin{equation} \label{eq:p3discrete-nonlinear}
\implies \bar{f}^h(\vec{x}, \vec{u}) = \bar{c}_0 \vec{x}^{[0]} + \bar{c}_1 \vec{x}^{[1]} =  (1 - a)^{-1}(\bar{f}(\vec{x}, \vec{u}) - a\vec{x}) \text{.}
\end{equation}
Note that (\ref{eq:p3discrete-nonlinear}) is consistent with (\ref{eq:p3discrete}).

\paragraph{Continuous Double Exponential Synapse}

For the double exponential synapse:
\begin{equation} \label{eq:double-exp}
H(s) = \frac{1}{(\tau_1 s + 1)(\tau_2 s + 1)} = \frac{1}{\tau_1 \tau_2 s^2 + (\tau_1 + \tau_2)s + 1}
\end{equation}
\begin{align} \label{eq:double-exp-solution}
\implies f^h(\vec{x}, \vec{u}) &= \vec{x} + (\tau_1 + \tau_2) \dot{\vec{x}} + \tau_1 \tau_2 \ddot{\vec{x}} \nonumber \\
&= \vec{x} + (\tau_1 + \tau_2) f(\vec{x}, \vec{u}) + \tau_1 \tau_2 \left( \frac{\partial f(\vec{x}, \vec{u})}{\partial \vec{x}} \cdot f(\vec{x}, \vec{u}) + \frac{\partial f(\vec{x}, \vec{u})}{\partial \vec{u}} \cdot \dot{\vec{u}} \right) \text{.}
\end{align}
In the linear case, this simplifies to:
\begin{align*}
f^h(\vec{x}, \vec{u}) = \left( \tau_1 \tau_2 A^2 + (\tau_1 + \tau_2) A + I \right) \vec{x} + \left( \tau_1 + \tau_2 + \tau_1 \tau_2 A \right) B \vec{u} + \tau_1 \tau_2 B \dot{\vec{u}} \text{.}
\end{align*}

\paragraph{Linear Systems}

As in \S\ref{sec:lti}, Theorems~\ref{thm:p3cont-nonlinear} and~\ref{thm:p3disc-nonlinear} require that we differentiate the desired dynamical system.
For the case of nonlinear systems, this means determining the (possibly higher-order) Jacobian(s) of $f$, as shown in (\ref{eq:double-exp-solution}).
For the special case of LTI systems, we can determine this analytically to obtain a closed-form expression.
By induction it can be shown that:
$$\vec{x}^{(i)} = A^i\vec{x} + \sum_{j=0}^{i-1} A^{i-j-1} B \vec{u}^{(j)} \text{.}$$
Then by expanding and rewriting the summations:
\begin{align}
f^h(\vec{x}, \vec{u}) &= \sum_{i=0}^k c_i \vec{x}^{(i)} \nonumber \\
&= \sum_{i=0}^k c_i \left[ A^i \vec{x} + \sum_{j=0}^{i-1} A^{i-j-1} B \vec{u}^{(j)} \right] \nonumber \\
&= \underbrace{\left( \sum_{i=0}^k c_i A^i \right)}_{\parbox{4em}{Recurrent
Matrix}}\vec{x} + \sum_{j=0}^{k-1} \underbrace{\left( \sum_{i=j+1}^k c_i A^{i-j-1} \right)B}_{\parbox{7em}{Input Matrices}} \vec{u}^{(j)} \text{.} \label{eq:p3general-lti}
\end{align}
The discrete case is identical:
\begin{align}
\bar{f}^h(\vec{x}, \vec{u}) = \underbrace{\left( \sum_{i=0}^k \bar{c}_i \bar{A}^i \right)}_{\parbox{4em}{Recurrent
Matrix}}\vec{x} + \sum_{j=0}^{k-1} \underbrace{\left( \sum_{i=j+1}^k \bar{c}_i \bar{A}^{i-j-1} \right)\bar{B}}_{\parbox{7em}{Input Matrices}} \vec{u}^{[j]} \text{.} \label{eq:p3general-lti-discrete}
\end{align}
This gives a matrix form for any LTI system with a $k^\text{th}$ order synapse, provided we can determine $\vec{u}^{(j)}$ or $\vec{u}^{[j]}$ for $0 \le j \le k-1$.
Again, (\ref{eq:p3general-lti}) is consistent with (\ref{eq:general-linear}) and (\ref{eq:general-linear-approx}), as is (\ref{eq:p3general-lti-discrete}) with (\ref{eq:general-linear-approx-discrete}).


\section{Accounting for Synaptic Heterogeneity}
\label{sec:p3-hetero}

We now show how \S\ref{sec:principle3} can be applied to train efficient networks where the $i^\text{th}$ neuron has a distinct synaptic filter $h_{i}(t)$, given by:
\begin{equation} \label{eq:synapse-general}
H_i(s) = \frac{1}{\sum_{j=0}^{k_i} c_{ij} s^j} \text{.}
\end{equation}
This network architecture can be modeled in Nengo using nengolib~0.4.0~\citep{nengolib}. 
This is particularly useful for applications to neuromorphic hardware, where transistor mismatch can change the effective time-constant(s) of each synapse.
To this end, we abstract the approach taken in \citep{voelker2017a}.
We show this specifically for Theorem~\ref{thm:p3cont-nonlinear}, but this naturally applies to all methods in this report.

Recalling the intuition behind Principle~III, our approach is to separately drive each synapse $h_{i}$ with the required signal $f^{h_i}(\vec{x}, \vec{u})$ such that each PSC becomes the desired representation $\vec{x}$.
Thus, the connection weights to the $i^\text{th}$ neuron should be determined by solving the decoder optimization problems for $f^{h_i}(\vec{x}, \vec{u})$ using the methods of \S\ref{sec:principle3} with respect to the synapse model $h_i$. 
This can be repeated for each synapse to obtain a full set of connection weights.
While correct in theory, this approach displays two shortcomings in practice: (1) we must solve $n$ optimization problems, where $n$ is the number of post-synaptic neurons, and (2) there are $\mathcal{O}(n^2)$ weights, which eliminates the space and time efficiency of using factorized weight matrices~\cite{mundy2015efficient}.

We can solve both issues simultaneously by taking advantage of the linear structure within $f^{h_i}$ that is shared between all $h_i$.
Considering Theorem~\ref{thm:p3cont-nonlinear}, we need to drive the $i^\text{th}$ synapse with the function:
\begin{equation*}
f^{h_i}(\vec{x}, \vec{u}) = \sum_{j=0}^{k_i} c_{ij} \vec{x}^{(j)} \text{.}
\end{equation*}
Let $\vec{d}^j$ be the set of decoders optimized to approximate $\vec{x}^{(j)}$, for all $j = 0 \ldots k$, where $k = \max_i k_i$.
By linearity, the optimal decoders used to represent each $f^{h_i}(\vec{x}, \vec{u})$ may be decomposed as:
\begin{equation*}
\vec{d}^{f^{h_i}} = \sum_{j=0}^{k_i} c_{ij} \vec{d}^j \text{.}
\end{equation*}
Next, we express our estimate of each variable $\vec{x}^{(j)}$ using the same activity vector $\vec{a}$:
\begin{align*}
\vec{\hat{x}}^{(j)} = \dotp{\vec{a}}{\vec{d}^j}{} \text{.}
\end{align*}
Now, putting this all together, we obtain:
\begin{equation} \label{eq:hetero-decoding}
\dotp{\vec{a}}{\vec{d}^{f^{h_i}}}{} = \sum_{j=0}^{k_i} c_{ij} \dotp{\vec{a}}{\vec{d}^j}{} = \sum_{j=0}^{k_i} c_{ij} \vec{\hat{x}}^{(j)} \approx f^{h_i}(\vec{x}, \vec{u}) \text{.}
\end{equation}

Therefore, we only need to solve $k+1$ optimization problems, decode the ``matrix representation'' $\left[ \vec{\hat{x}}^{(0)}, \vec{\hat{x}}^{(1)}, \ldots, \vec{\hat{x}}^{(k)} \right]$, and then linearly combine these $k+1$ different decodings as shown in (\ref{eq:hetero-decoding})---using the matrix of coefficients $c_{ij}$---to determine the input to each synapse.
This approach reclaims the advantages of using factorized connection weight matrices, at the expense of a factor $\mathcal{O}(k)$ increase in space and time efficiency.

\section{Discussion}

We have reviewed three major extensions to the NEF that appear in recent publications.
These methods can be used to implement linear and nonlinear systems in spiking neurons recurrently coupled with heterogeneous higher-order mixed-analog-digital synapses.
This provides us with the ability to implement NEF networks in state-of-the-art neuromorphics while accounting for, and sometimes even exploiting, their nonideal nature.

While the linear and nonlinear methods can both be used to harness pure spike time-delays (due to axonal transmission) by modeling them in the synapse, the linear approach provides greater flexibility.
Both extensions can first transform the time-delay into the standard form of (\ref{eq:synapse}) via Pad\'e approximants, which maintains the same internal representation $\vec{x}$ as the desired dynamics (within some radius of convergence).
But the linear extension also allows the representation to change, since it is only concerned with maintaining the overall input-output transfer function relation.
In particular, we derived an analytic solution using the Lambert-$W$ function, which allows the neural representation $\vec{x}$, and even its dimensionality, to change according to the expansion of some Taylor series.
The linear case is also much simpler to analyze in terms of the network-level transfer function that results from substituting one synapse model for another.
For all other results that we have shown, the linear extension is consistent with the nonlinear extension, as they both maintain the desired representation by fully accounting for the dynamics in the synapse.

\section*{Acknowledgements}

We thank Wilten Nicola for inspiring our derivation in \S\ref{sec:nonlinear} with his own phase-space derivation of Principle~III using double exponential synapses for autonomous systems (unpublished).
We also thank Kwabena Boahen and Terrence C. Stewart for providing the idea used in \S\ref{sec:p3-hetero} to separately drive each $h_i$, and for improving this report through many helpful discussions.

\newpage

\bibliographystyle{ieeetr}
{\footnotesize \bibliography{refs}}

\end{document}